\documentstyle[aps,preprint]{revtex}
\tightenlines
\begin{document}

\pagestyle{plain}
\setcounter{page}{1}
\setcounter{footnote}{00}

\renewcommand{\thefootnote}{\alph{footnote}}

\baselineskip=18pt 
\def\doublespaced{\baselineskip=\normalbaselineskip\multiply
    \baselineskip by 150\divide\baselineskip by 100}


%
%
\begin{titlepage}
\baselineskip=0.2in
\begin{flushright}
HD-THEP-98-21\\
TUIMP-TH-98/96\\
\end{flushright}
\vspace{0.2in}
\begin{center}
{\large 
  CP Violation in  Top Quark Pair Production at Hadron Colliders }\\
\vspace{.2in}
Hong-Yi Zhou\\
\vspace{.2in}
CCAST (World Laboratory), 
\hspace{.2cm}
      P. O.\hspace{0.2cm}  Box 8730, Beijing 100080, P.R. China  and \\
        Institut f\"{u}r Theoretische  Physik, Universit\"{a}t Heidelberg,
        Philosophenweg 16, D-69120 Heidelberg, Germany
{\footnote{Present mailing address}}\\
              and \\
    Institute of Modern Physics and Department of Physics,\\
    Tsinghua University, Beijing 100084, P.R. China\\

\end{center}
\vspace{.3in}

\begin{center}\begin{minipage}{5in}
\baselineskip=0.25in
\begin{center} Abstract\end{center}

CP violating effects in top quark pair production at the future 
2 TeV $p\bar p$ Tevatron and 14 TeV $pp$ LHC colliders are investigated. 
We study three kinds of  CP violating sources:
the supersymmetric CP-odd phase of the stop 
trilinear soft breaking term: $arg(A_t)$, the CP-odd parameter in 
two-Higgs doublet extensions of the standard model(2HDM), 
and the model-independent top quark chromoelectric dipole moment(CEDM), 
respectively. Optimal observables as well as simple observables are used. 
We find that it is possible to observe CP violating effects from 
$arg(A_t)$ in top quark pair production at the 2 TeV Tevatron 
with $\sim 30fb^{-1}$ integrated luminosity 
when $m_{\tilde{g}}\sim 200$ GeV. If the experimental 
systematic errors are sufficient small, the LHC  with $\sim 150fb^{-1}$ 
can put a limit of order $10^{-1}$ on the phase $arg(A_t)$ 
and the CP-odd parameter in 2HDM by using optimal observables.  
The CEDM of the top quark can be measured to an accuracy of 
$10^{-18} ~cm~g_s$ at the Tevatron 
and few $\times 10^{-20} ~cm~g_s$ at the LHC.

\end{minipage}\end{center}

\vspace{.5in}

PACS number(s): 11.30.Er, 14.65.Ha, 12.38.Bx

\end{titlepage}
\newpage

\baselineskip=18pt 
\renewcommand{\thefootnote}{\arabic{footnote}}
\setcounter{footnote}{0}

\section{ Introduction }
\indent

At the future hadron colliders such as the upgraded Tevatron with 
$\sqrt{s}=2$ TeV and the CERN LHC with $\sqrt{s}=14$ TeV, the 
annual top quark pair yields are about $7\times 10^4$
(with integrated luminosity $\int {\cal L}=10 fb^{-1}$) and $7\times10^7$ 
(with  $\int {\cal L}=100 fb^{-1}$), respectively. 
The large numbers of top quark pairs allow us to do precise 
measurements on physical quantities associated with the top quark. 
Among them are the production 
cross sections, the top quark mass and the discrete symmetry 
properties.  Due to the QCD uncertainties  and the experimental 
systematic errors, the precision of cross section measurement is only 
about $5\%-6\%$\cite{tev200}. The discrete symmetry properties such 
as parity nonconservation and CP violation do not suffer from 
QCD background uncertainty and their accuracies of measurements 
depends mainly on the statistical errors provided the experimental 
systematic errors are sufficient small. 
Therefore the discrete 
symmetry properties can be measured more precisely than the 
cross sections. New physics which has no observable contributions 
to cross section may have observable effects in parity or CP violation. 
Since the Standard Model(SM) contributions to parity violation\cite{hollik} 
and CP violation are small, possible observed large 
effect of them will reveal new physics.

In a previous work\cite{pvtv}, we have shown that the minimal 
supersymmetric(SUSY) extension of the standard model (MSSM)\cite{haber} gives 
observable parity violating effects at the Tevatron while its 
corrections to the production cross section  are within the QCD 
theorectical uncertainties. In this paper, we shall concentrate on the 
CP violating effects induced by non-standard model interactions. 
We investigate three kinds of  CP violating sources:
the supersymmetric CP-odd phase of the stop 
trilinear soft breaking term:$arg(A_t)$, the CP-odd parameter in 
two-Higgs doublet extensions of the standard model(2HDM), 
and the model-independent top quark chromoelectric dipole moment(CEDM),
respectively. 
In the MSSM, CP violation exists in strong interaction and 
in the 2HDM, CP violation can have strong Yukawa  couplings due to the heavy 
top quark mass. Therefore, both can produce possible large  effects. It 
is also useful to study CP violation in a model-independent way when 
we do not know what is  the new physics.  
CP violation in top quark pair production at hadron colliders 
is also studied in Refs.\cite{cpyuan}-\cite{polpp}. 
In Ref.\cite{schmidt}, SUSY QCD 
CP violating effects are studied in the $gg\to t\bar t$ process 
by using a charge energy asymmetry observable which is only 
sensitive to the imaginary part of the loop integrals. 
In this work, we use the optimal observables  as well as naive 
observables constructed from the final state momenta. 
Possible large CP violating effects in 2HDM 
and the methods of observing 
them in  top quark pair production at hadron colliders are 
studied in Refs.\cite{peskin}\cite{bern}. We extend those 
studies by applying optimal observables. 
In Ref.\cite{atwood1}, the method of extracting real top 
quark  CEDM 
in the reaction $gg\to t\bar t$ is studied. It is found that the optimal 
observables are particularly effective. We include here an imaginary 
part of CEDM and the reaction $q\bar q\to t\bar t$ at the Tevatron. 
Furthermore, we use  the exact amplitudes of $gg(q\bar q)\to t\bar t
\to bl_1^+\nu_{l_1}\bar b l_2^-{\bar\nu}_{l_2}$ 
( $b\bar q_1 q_1' \bar b q_2\bar q'_2$). 
In Ref.\cite{atwood1}, the top quark spin in its rest frame 
is taken to be in the direction of the lepton. This is a kind 
of approximation, althogh the lepton is a good analyzer of the top 
quark spin, because in the top rest frame, the lepton momentum 
has the angular distribution proportional 
to $\displaystyle 1+\cos\psi$ with $\psi$ being the angle 
between the top spin and the lepton momentum. 
The possibility of using polarized 
proton was studied in Ref.\cite{polpp}.

In Sec II, we describe  the  models and the calculations. 
The methods of extracting CP violating effects are given in 
Sec. III.  In Sec IV, we present our results, discussions 
and conclusions.

\section{ Models and Calculations }
\indent

\begin{flushleft} 
{\bf A. CP violation in MSSM} 
\end{flushleft}

In the MSSM, two possibilities can induce CP violation in top quark 
interactions: the complex phase in Higgs mass parameter $\mu$, 
and the complex phase in scalar top supersymmetric soft breaking 
trilinear coupling $ A_t$. The experimental limit on the Neutron Electric 
Diploe Moment(NEDM), $d_n\leq 1.1\times 10^{-25} e-cm$\cite{nedm},
places a severe constraint on the phase of $\mu$. Therefore, the 
only significant SUSY CP-odd phase in associate with the top quark 
is $arg(A_t)$. In Ref.\cite{atwood2}, it is argued that the phase 
$arg(A_t)$ is not strongly constrained by current experiments and 
the effects in single top quark production and decay are studied.  
In this work, 
we shall assume  $arg(\mu)=0$ and let $arg(A_t)$ to be a free 
parameter of no {\it a priori} constraints. 
 
The parameter $arg(A_t)$ enters in the scalar top quark mixing.
The mass eigenstates $\tilde{t}_1$ and
$\tilde{t}_2$ of scalar top quark are related to the current 
eigenstates $\tilde{t}_L$ and $\tilde{t}_R$ by 
\begin{equation}
\tilde{t} _1 = \tilde{t} _L \cos \theta _t + \tilde{t} _R \sin \theta 
_t e^{-i\beta_t},~~~\tilde{t} _2 = -\tilde{t} _L \sin \theta _t 
e^{i\beta_t} + \tilde{t} _R \cos \theta _t 
\end{equation}  

The mixing angle $\theta_t$, phase $\beta_t$ as well as the masses 
$m_{\tilde{t}_{1,2}}$
can be calculated by diagonalizing the following mass matrix\cite{haber}

\begin{eqnarray}
\label{eqnum2}
& &  M^2_{\tilde{t}} =\left(
           \begin{array} {ll}
             M^2_{\tilde{t}_L} & m_tm_{LR}^\star \\
             m_tm_{LR} & M^2_{\tilde{t}_R} 
           \end{array}
         \right)~~,    \nonumber \\
& & M^2_{\tilde{t}_L}=m^2_{\tilde{t}_L} + m^2_t+(\frac{1}{2}
  -\frac{2}{3}\sin^2\theta_W)\cos(2\beta)m_Z^2\nonumber~~, \\
& &  M^2_{\tilde{t}_R}=m^2_{\tilde{t}_R} + m^2_t
  +\frac{2}{3}\sin^2\theta_W\cos(2\beta)m_Z^2\nonumber~~, \\
& & m_{LR}=-\mu\cot\beta-A_t~~,
\end{eqnarray}   
where $m^2_{\tilde{t}_L},~m^2_{\tilde{t}_R}$ are the soft SUSY-breaking 
mass terms of left- and right-handed stops,  
$\tan\beta=v_2/v_1$ is the ratio of the vacuum expectation values 
of the two Higgs doublets. 

From Eqs.(1)(2), we can get the expressions for 
$m^2_{\tilde{t}_{1,2}}$, $\theta_t$ and $\beta_t$ : 
\begin{eqnarray}  
& & m^2_{\tilde{t}_{1,2}}=\frac{1}{2}\left[ 
M^2_{\tilde{t}_{L}}+M^2_{\tilde{t}_{R}}\mp\sqrt{(M^2_{\tilde{t}_{L}}
-M^2_{\tilde{t}_{R}})^2+4m_t^2|m_{LR}|^2}\right]~~,\\
& &\tan\theta_t=\frac{m^2_{\tilde{t}_1}-M^2_{\tilde{t}_{L}}}
{m_t(-\mu\cot\beta-|A_t|\cos\theta_{A_t})}\cos\beta_t~~,\\
& &\tan\beta_t=\frac{|A_t|\sin\theta_{A_t}}{\mu\cot\beta +|A_t|\cos\theta
_{A_t}}~~,
\end{eqnarray}
where $\theta_{A_t}=arg(A_t)$ .

In the presence of squark mixing,the strong squark-quark-gluino interaction 
Lagrangian is given by\\
\begin{eqnarray}
 L_{\tilde{g} \tilde{q} \bar{q}} & = & -g_s T^a_{jk}\bar{q}_k[(a_1-b_1
\gamma_5)\tilde{q}_{1j}+(a_2-b_2\gamma_5)\tilde{q}_{2j}]\tilde{g}_a+H.C.\;,
\end{eqnarray}          
where $g_s$ is the strong coupling constant , $T^a$ 
are $SU(3)_C$ generators and for top quark 
$a_1,~b_1,~a_2,~b_2$ are given by
\begin{eqnarray}
& & a_1=\frac{1}{\sqrt{2}}(\cos\theta _t-\sin\theta_t e^{i\beta_t}),\;\; 
 b_1=-\frac{1}{\sqrt{2}}(\cos\theta _t+\sin\theta_t e^{i\beta_t}),\nonumber\\
& & a_2=-\frac{1}{\sqrt{2}}(\cos\theta _t+\sin\theta_t e^{-i\beta_t}),\;\;
 b_2=-\frac{1}{\sqrt{2}}(\cos\theta _t-\sin\theta_t e^{-i\beta_t}).
\end{eqnarray}  

The above interactions enter in the virtual corrections to the main 
production processes of $ t\bar t$ at hadron colliders: $q\bar q\to t\bar t$ 
and $gg\to t\bar t$. There are also weak squark-quark-neutralino and 
squark-quark-chargino interactions. Since their coupling constants 
are an order of magnitude smaller than the strong SUSY QCD 
squark-quark-gluino interaction, we shall not consider them here. 
However, they may be important when the main processes are weak 
interactions\cite{atwood2}.

In Fig.1(a)--(e), the Feynman diagrams of 
the QCD tree level and SUSY QCD  virtual corrections of the process 
$ q\bar q\to t\bar t$ are given. The corresponding Feynman diagrams 
of $gg\to t\bar t$ are presented in Fig.2(a)--(m) (the u-channels of 
Fig.2 (b),(f)--(m) are not depicted). 
The dashed lines in the loop 
stand for scalar quarks, while the solid lines for gluinos.
In Ref.\cite{schmidt},
CP violating effects are studied using a 
charge energy asymmetry observable which is only 
sensitive to the imaginary part of the loop integrals. 
Therefore,  there are no contributions to the charge energy 
asymmetry from Fig.2 (f)--(j),(m).

The one loop scattering
amplitudes of $ q \bar q \to t \bar t$ and $ gg \to t \bar t$ 
were already presented 
in Refs. \cite{rateone,ratetwo,zhy1} for calculating the 
total production rates of $ t \bar t$ pairs. 
To calculate the CP violating effects in the  $ t \bar t$ system, 
additional renormalized amplitudes are needed.
In terms of the tree-level amplitude, $M_{0}^{a}$, and the next-to-leading
order SUSY QCD corrections, $\delta M^{a}$, the renormalized amplitudes 
of $a\bar a\to t\bar t$ ($a=q,~g$) at the one-loop level may be written as
\begin{eqnarray}
M^a=M_0^a+\delta M^a.
\end{eqnarray}
where $\delta M^a$ can be decomposed into two parts: $\delta M^{aS}$ which 
contains even  combination of  $\gamma_5$ 
and $\epsilon_{\mu\nu\rho\sigma}$, 
and  $\delta M^{aA}$ containing odd combination of  $\gamma_5$
and $\epsilon_{\mu\nu\rho\sigma}$. 
The symmetry breaking effects are contained in
$\delta M^{aA}$ which has no contributions to the 
total cross sections at next-to-leading order, 
while $\delta M^{aS}$ will contribute to the total 
cross sections. We shall assume that $\delta M^{aS}$ is  
small enough to be within the $5\%-6\%$ uncertainty therefore is 
neglected in our calculations. We also discard terms in $\delta M^{aA}$ 
which give only parity asymmetry\cite{pvtv}.

Let us denote the momenta of the initial and the final state
particles as $ a(p_4)\bar a(p_3)\to t_i(p_2)\bar t_j(p_1)$.
We may use, as a further short-hand, the notation for $a=q$ that 
$u_{i} \equiv u(p_{i})$ ($v_i \equiv v(p_i)$)
denotes the Dirac four-spinor corresponding to
the momentum and spin of particle (anti-particle). When $a=g$ 
we use   $\epsilon_i\equiv \epsilon(p_i)$ for the gluon 
polarization function. 
In this notation,
the tree level amplitude for $a=q$ can be written as
\begin{equation}                          
M_0^q=ig_s^2(T^c_{ji}T^c_{lm})\bar v_3\gamma^\mu u_4
\bar u_2\gamma_\mu v_1/\hat{s},
\end{equation}        
where $\hat s$ is the invariant mass of the $t \bar t$ pair.

The tree level amplitude for $a=g$ is composed of three different 
production channels(s-,t-,u-channel) as following:

\begin{eqnarray}                                                
 M_0^{gs} &= & -ig_s^2(if_{abc}T^c)_{ji}{\bar u}(p_2)
 \rlap/\Gamma v(p_1)/\hat{s}\nonumber\\
&=&-ig_s^2(T^aT^b-T^bT^a)_{ji}M_0^{s},  \\
 M_0^{gt}& = & -ig_s^2(T^bT^a)_{ji}{\bar u}(p_2)\rlap/\epsilon_4
 (\rlap/q+m_t)\rlap/\epsilon_3 v(p_1)/(\hat{t}-m_t^2)\nonumber\\
& =& -ig_s^2(T^bT^a)_{ji}M_0^{t},\\
 M_0^{gu}& = & M_0^{gt}(p_3\leftrightarrow p_4,\;\;T^a\leftrightarrow T^b,\;\;
 \hat{t}\rightarrow \hat{u} )\nonumber\\
&=&-ig_s^2(T^aT^b)_{ji}M_0^{u},
\end{eqnarray}
where  $q=p_2-p_4$, $\Gamma^\mu$ is given in the Appendix A.

To calculate the CP violating effects 
induced by the SUSY QCD effects,
we follow the method presented in 
Ref. \cite{cpyuan}\cite{ZEPPEN}, in which the amplitudes were calculated 
numerically using the helicity amplitude method.
To obtain the renormalized scattering amplitudes, we adopt the
dimensional regularization scheme to regulate 
the ultraviolet divergences and the on-mass-shell 
renormalization scheme to subtract the divergences\cite{onshell}.

The SUSY QCD corrections to the scattering amplitude $\delta M^{aA}$ 
for $a=q$ arise from the vertex diagram, the box diagram, as well as the
crossed-box diagram. 
The renormalized amplitudes can be written as
\begin{eqnarray}
\delta M^{qA} & =& \delta M^{qv}+\delta M^{DB}+\delta M^{CB},
\end{eqnarray}
where $\delta M^{qv}$ is the vertex corrections
of Fig.1 (b)(c), and $\delta M^{DB}$
and $\delta M^{CB}$ are the contributions from the box diagram and
crossed-box diagram of Fig.1 (d)(e), respectively.
The results for these separate contributions are,
\begin{eqnarray}                      
\label{qqv}
\delta M^{qv}& =& ig_s^2 (T^c_{ji}T^c_{lm})\bar v_3(\rlap/p_2-\rlap/p_1)u_4 
{\bar u}_2(-D^s\gamma_5)v_1/\hat{s},
\end{eqnarray}
\begin{eqnarray}                      
\delta M^{DB}& =&ig_s^2\frac{7}{6}(T_{ji}^cT_{lm}^c)f^{DB}[
\bar u_2\rlap/p_3 \gamma_5u_4\bar v_3v_1-\bar u_2\rlap/p_3
u_4\bar v_3\gamma_5v_1\nonumber \\
& &+\bar u_2 u_4\bar v_3\rlap/p_4\gamma_5 v_1
+\bar u_2\gamma_5 u_4\bar v_3\rlap/p_4v_1],
\end{eqnarray}
\begin{eqnarray}                      
\delta M^{CB}& =& \frac{2}{7} \delta M^{DB}(p_3\leftrightarrow p_4).  
\end{eqnarray}
The form factor $D^s$ corresponding to the top quark 
CEDM and the form factor $f^{DB}$ 
are given in the Appendix A.

The SUSY QCD corrections to the scattering amplitude $\delta M^{gA}$ 
of $gg\to t\bar t$ can be written as

\begin{eqnarray}
& & \delta M_{ji}^{gA}=-ig_s^2[\delta M^+  O^{(+)}_{ji}
+\delta M^-O^{(-)}_{ji}                   
+\delta M^{\delta}[\delta_{ab}]_{ji}]\;\;, 
\end{eqnarray} 
with 
\begin{eqnarray} 
& & O^{(+)}=\frac{T^aT^b+T^bT^a}{2}, \;\;\;
O^{(-)}=\frac{T^bT^a-T^aT^b}{2},
\end{eqnarray}

\begin{eqnarray} 
\delta M^+&=&-\frac{1}{3}\delta M^{s2}+\delta M^{self,t}
+\delta M^{self,u}+\delta M^{v1,t}+\delta M^{v1,u}
+\delta M^{v2,t}  \nonumber\\
& &+\delta M^{v2,u}+\delta M^{box1,t}
+\delta M^{box1,u}
+\delta M^{box2,t}+\delta M^{box2,u}\nonumber\\
\delta M^-&=&-2\delta M^{s1}
+\delta M^{self,t}-\delta M^{self,u}
+\delta M^{v1,t}-\delta M^{v1,u}\nonumber\\
& & +\delta M^{v2,t}-\delta M^{v2,u}  
+\delta M^{box1,t}-\delta M^{box1,u}
+\delta M^{box2,t}-\delta M^{box2,u}\nonumber\\
\delta M^{\delta}&=&\frac{1}{2}\delta M^{s2}
+\frac{1}{6}(\delta M^{box1,t}+\delta M^{box1,u})
-\frac{3}{2}(\delta M^{box2,t}+\delta M^{box2,u})\nonumber\\
& &-(\delta M^{box3,t}+\delta M^{box3,u})
\end{eqnarray}                   

In the above expressions, the superscript $t,u$ stand for 
t-,u-channel.   
$\delta M^{s1}$ is the s-channel vertex
corrections from Fig.2 (c)(d), $\delta M^{s2}$ from (e),  
$\delta M^{self,t}$ the self energy correction from (f), 
$\delta M^{v1,t}$ from (g)(h),$\delta M^{v2,t}$ from (i)(j),
$\delta M^{box1,t}$ from (k), $\delta M^{box2,t}$ from (l) 
and $\delta M^{box3,t}$ from (m).  
In the following we only give the explicit results of the s-channel
(no crossed diagram) and t-channel contributions. 
The u-channel results can be obtained by the 
following substitutions:
\begin{eqnarray}                               
& & p_3\leftrightarrow p_4,\;\;
a\leftrightarrow b,\;\;
\hat{t}\leftrightarrow\hat{u}.
\end{eqnarray}        
All s-, t-channel 
terms in $\delta M^{+},~~\delta M^{-},~~\delta M^{\delta}~~$ 
can be written as following according to their  
Lorentz structures: 
\begin{eqnarray}                      
\label{mx}              
\delta M^X & =& \epsilon^{\mu}_4\epsilon^{\nu}_3
\bar{u}_2[f_1^X g_{\mu \nu}+f_2^X \gamma_{\mu} \gamma_{\nu} 
+f_7^Xp_{1\mu}p_{1\nu} +f_8^Xp_{1\mu}p_{2\nu}
+ f_9^X p_{2\mu}p_{1\nu}\\ \nonumber  
& & + f_{10}^Xp_{2\mu}p_{2\nu}
+ f_{13}^X\rlap/p_4 p_{1\nu}\gamma_{\mu} + f_{14}^X \rlap/p_4 p_{1\mu}
\gamma_{\nu}+ f_{15}^X\rlap/p_4 p_{2\nu}\gamma_{\mu}
+ f_{16}^X\rlap/p_4 p_{2\mu} \gamma_{\nu}]\gamma_5v_1\,,
\end{eqnarray}
where $X=s1,~s2,~self,~v1,~v2,~box1,~box2,~box3$, respectively. 
The 10 form factors corresponding to each diagram are given in 
the Appendix A. They are not all independent when we sum over all 
possible channels. CP-odd property of $\delta M^X$ requires
\begin{eqnarray}
f_{13}^X=f_{16}^X~,~~~f_{14}^X=f_{15}^X~,~~~f_{7}^X+2f_{14}^X=f_{10}^X~. 
\end{eqnarray}
Those  relations are verified by our explicity formula of the form 
factors. We find that all form factors are proportional to 
$\lambda_{CP}=2Im(a_1b_1^\star)=\sin2\theta_t\sin\beta_t$ .

The color sum of the  amplitude square including the next-to-leading 
order correction is:
\begin{eqnarray}  
\label{colorsum}
& &\sum\limits_{color}\{|M_0^g|^2+2Re(M_0^g\delta M^{gA\dagger})\}
\\ \nonumber
& & = g_s^4\{\frac{7}{3}|M_0^+|^2+3|M_0^-|^2\nonumber\\\nonumber
& &+\frac{14}{3}Re(M_0^+\delta M^{+\dagger})+6Re(M_0^-\delta M^{-\dagger})
+8Re(M_0^+\delta M^{\delta\dagger})\}~,
\end{eqnarray}
where 
\begin{eqnarray} 
& & M_0^+=M_0^{t}+M_0^{u},\;\;\;
M_0^-=M_0^{t}-M_0^{u}-2M_0^{s},
\end{eqnarray}

\begin{flushleft} 
{\bf B. CP violation in 2HDM} 
\end{flushleft}

In ordinary 2HDM, there are 
three neutral physical Higgs bosons, namely two CP-even scalars $H,~h$, 
one CP-odd pseudoscalar $A$. CP violation in the scalar potential\cite{cppot} 
induces mixing of the CP-even and CP-odd neutral Higgs bosons, 
thus leading to three physical mass eigenstates $|\phi_j>~(j=1,2,3)$ 
with no definite CP parity. Their Yukawa couplings to the top quark 
can be written as(in the notation of \cite{ntt})
\begin{eqnarray}
L_Y&=&-m_t(\sqrt{2}G_F)^{1/2}\sum\limits_{j=1}\limits^{3} 
\bar t (a_{jt}+i\tilde{a}_{jt}\gamma_5)t \phi_j
\end{eqnarray} 
where $G_F$ is the Fermi constant, and 
\begin{eqnarray} 
& & a_{jt}=d_{2j}/\sin\beta~,~~\tilde{a}_{jt}=-d_{3j}\cot\beta~,
\end{eqnarray} 
$d_{2j}~,d_{3j}$ are the matrix elements of a $3\times 3$ 
orthogonal matrix which describes the mixing of the neutral states\cite{ntt}.
We  assume that the two heavier Higgs bosons($j=2,3$) may be neglected and 
define $a=a_{1t},~\tilde{a}=\tilde{a}_{1t},~\phi=\phi_1$. 
The strength of CP violation is proportional to $2Im[a_{1t}(-i\tilde{a}
_{1t})^\star]=2a\tilde{a}=-2\gamma_{CP}$($\gamma_{CP}$ is 
defined in \cite{bern}).

The one-loop Feynman diagrams of Higgs boson contributions to 
the processes $q\bar q\to t\bar t$ and $gg\to t\bar t$ can be 
represented by Fig.1(b) and Fig.2(c)(f)(g)(i)(k)(n), respectively. 
Now the dashed lines in the loop 
stand for Higgs bosons, while the solid lines for top quarks. 

The one-loop amplitudes can be easily obtained from those of the 
MSSM by the method given in the Appendix A with no changes on  
$\delta M^{qv}~,~~\delta M^+~,~~\delta M^-$ and the modification 
on $\delta M^{\delta}$ which comes from Fig.2 (n):
\begin{eqnarray}
\label{res}
\delta M^{\delta}
&=& \epsilon_4^\mu\epsilon_3^\nu \bar u_2\{
a^2(f^{sr}_1g_{\mu\nu}+f_2^{sr}p_{3\mu}p_{4\nu})
-\tilde{a}^2f_3^{sr}i\epsilon_{\mu\nu\rho\sigma}p_4^\rho p_3^\sigma 
\gamma_5\\ \nonumber
& &+a\tilde{a}[i(f_1^{sr}g_{\mu\nu}+f_2^{sr}p_{3\mu}p_{4\nu})
\gamma_5
-f_3^{sr}\epsilon_{\mu\nu\rho\sigma}
p_4^\rho p_3^\sigma]\}v_1/(\hat{s}-m_{\phi}^2+im_{\phi}\Gamma_\phi) 
\end{eqnarray} 
where $\Gamma_\phi$ is the Higgs boson width and $f^{sr}_{1,2,3}$ 
are given in the Appendix A. $\delta M^{\delta}$ is composed of two parts: 
$\delta M^{\delta e}$, which is CP-even and contains terms of 
$a^2$, $\tilde{a}^2$; $\delta M^{\delta o}$, the CP-odd term proportional 
to $a\tilde{a}$. We keep the CP-even term because when $m_{\phi}>2m_t$,  
it may be important. The $8Re(M_0^+\delta M^{\delta\dagger})$
term in the 
color sum of (\ref{colorsum}) should be replaced by   
\begin{eqnarray}
\label{twohd}
8Re(M_0^+\delta M^{\delta\dagger})
+24(|\delta M^{\delta e}|^2+|\delta M^{\delta o}|^2)+
48 Re(\delta M^{\delta e}\delta M^{\delta o\dagger}) ~.
\end{eqnarray}
We compute $\Gamma_\phi$ by using the formula and parameters 
of Ref.\cite{bern}
\footnote{The misprint of $\Gamma_Z$ is corrected.}.

\begin{flushleft} 
{\bf C. CP violation from model-independent top quark CEDM} 
\end{flushleft}

When the particles in the loops  are heavy compared to the 
external particle momenta, it is convenient to describe the 
loop induced interactions by effective interaction Lagrangian.
Even though when the loop masses are not too large, a model-independent 
study can give us knowledge about the sensitivities of given 
colliders.         
That approach is particularly useful when we do not know the 
underlying new physics. We assume the following additional 
top-quark-gluon effective interaction:
\begin{eqnarray}
L_D=\frac{id_t}{2}\bar t \sigma^{\mu\nu}\gamma_5 F_{\mu\nu}^a T^a t~,
\end{eqnarray}
where $\sigma^{\mu\nu}=(i/2)[\gamma^\mu,\gamma^\nu]$, $F_{\mu\nu}^a $
is the the gluon field strength. The coefficient $d_t$ is the 
top quark chromoelectric diploe moment(CEDM) which we assume to have  
imaginary part as well as real part. We denote it as $d_t^R+id_t^I$.

The CEDM $d_t$ contributes to the CP violating amplitudes through the 
diagrams of Fig.3.  Let us define $\hat{d}_t=d_t/g_s=\hat{d}_t^R
+i\hat{d}_t^I$. 
The contribution of  Fig.3 (a) to $q\bar q \to t\bar t$  
can  be obtained by simply replacing $D^s$ with $i\hat{d}_t$ in Eq.(\ref{qqv}) 
and setting $\delta M^{DB}=\delta M^{CB}=0$. 
We denote the contributions to 
$gg \to t\bar t$  of Fig.3(b)--(e) as 
$\delta M^s$, 
$\delta M^b$, $\delta M^{v,t}$, $\delta M^{v,u}$(for the crossed diagrams), 
respectively. Then we have 
\begin{eqnarray} 
\delta M^+&=& \delta M^{v,t}+\delta M^{v,u}~,\nonumber\\
\delta M^-&=&-2\delta M^{s}-2\delta M^{b} 
+\delta M^{v,t}-\delta M^{v,u}~,\nonumber\\
\delta M^{\delta}&=& 0~.
\end{eqnarray}                   
$\delta M^s$, 
$\delta M^b$, $\delta M^{v,t}$ can also be written as the form of 
Eq.(\ref{mx}) with the constant form factors given in the Appendix A. 
 
\section{Extracting the CP violating effects} 
\indent

Since $\delta M^{aA}$ contains odd combination of  $\gamma_5$
and $\epsilon_{\mu\nu\rho\sigma}$, its interference with the tree 
level amplitude will be zero if we sum up all the initial and final 
state spins or polarizations.  To see the CP violating effects, the spins 
of at least some of the particles must be identified. Because the initial 
state spins are difficult to be determined, the helicities of the final 
$t \bar t$ must be inferred from their decays. If we assume the 
SM couplings of the top quark to its decay products and 
set $m_b=0$, all decay products of the top quark will have left-handed 
helicities. That means the spin information of the top quark can 
only be transfered to the momentum correlations among the decay products. 

We assume the SM decay
of the top quark\footnote{The top quark decay asymmetry in the MSSM is 
found to be small\cite{atwood2}\cite{aoki}.} 
and apply the narrow width 
approximations of the top quark and W-boson propagators:
\begin{eqnarray}                  
\frac{1}{|q^2_Y-m_Y^2+im_Y\Gamma_Y|^2}\rightarrow 
\frac{\pi}{m_Y\Gamma_Y}\delta(q_Y^2-m_Y^2)~.
\end{eqnarray} 
where $Y$ stands for top quark and W-boson, $\Gamma_Y$ is the width of 
$Y$  
                  
The parton level cross section for reaction 
$a\bar a \to t\bar t \to bl_1^+\nu_{l_1}\bar b l_2^-{\bar\nu}_{l_2}$ 
($b\bar q_1 q_1' \bar b q_2\bar q'_2$)
can be written as:
\begin{eqnarray} 
\label{partonc}
d\hat{\sigma}_{a\bar a}&=& \frac{\gamma}{(8\pi)^{10}
\hat{s}}\frac{\lambda_t \overline{|M|}_{a}^2}
{m_t^2m_W^2\Gamma_t^2\Gamma_W^2}d\Omega_t d\Omega_{W^+}' d\Omega_{W^-}'
d\Omega_{l_1^+}'d\Omega_{l_2^-}'    
\end{eqnarray} 
where $\gamma=\sqrt{1-4m_t^2/\hat{s}}$ and 
\begin{eqnarray} 
\lambda_t=(1-\frac{(m_W+m_b)^2}
{m_t^2})(1-\frac{(m_W-m_b)^2}{m_t^2})\approx (m_t^2-m_W^2)^2/m_t^4~,
\end{eqnarray} 
$d\Omega_{W^+}' (d\Omega_{W^-}')$ is the solid angle element of $W^+(W^-)$
in the rest frame of the (anti) top quark,  
$d\Omega_{l_1^+}'(d\Omega_{l_2^-}')$ denotes the     
solid angle element of $l_1^+(l_2^-)$ in the rest frame of $W^+(W^-)$, 
$\overline{|M|}^2_a$ 
is the average amplitude  square excluding the top quark and W-boson 
propagators after the decays of the top quarks:  
\begin{eqnarray}  
\overline{|M|}^2_a
&=& S_a^{-1}\sum\limits_{color,spin}
\{|M_0^a|^2+2Re(M_0^a\delta M^{aA\dagger})\}~,
\end{eqnarray} 
where $1/S_a$ is the color,spin average factor: $S_q=36$ and $S_g=256$.
In our calculations, $\overline{|M|}^2_a$ is easily 
obtained from the amplitude of 
$a\bar a\to t\bar t$ by the following 
substitutions:
\begin{eqnarray} 
& & \bar u_2\rightarrow \frac{g^2}{8}\bar u_b \gamma_\mu(1-\gamma_5)
(\rlap/p_2+m_t)\bar u_{\nu_1}\gamma^\mu(1-\gamma_5) v_{l_1}~,\\\nonumber
& & v_1\rightarrow \frac{g^2}{8}\bar u_{l_2} \gamma_\mu(1-\gamma_5)
v_{\nu_2}(\rlap/p_1-m_t)\gamma^\mu(1-\gamma_5) v_{\bar b}~,
\end{eqnarray} 
where $g$ is the weak $SU(2)$ coupling constant. The above expresssions  
are calculated numerically.

The hadronic cross section is obtained by convoluting (\ref{partonc}) 
with parton distribution functions:
\begin{eqnarray}
\label{hadroncr} 
d\sigma_{p\bar p(p)}&=&\sum\limits_{a=q,g}\int_0^1 dx_1
\int_0^1 dx_2 [f_a^{p}(x_1)f_{\bar a}^{\bar p(p)}(x_2)+(
p\leftrightarrow \bar p(p))]d\hat{\sigma}_{a\bar a}/(1+\delta_{a,\bar a})~,
\end{eqnarray} 
where $f_a^{\bar p(p)}$, $f_{\bar a}^{\bar p(p)}$ are parton 
distribution functions(PDFs). 

It has been studied extensively in the literature
how to pick out the momentum correlation information contained 
in $2Re(M_0^a\delta M^{aA\dagger})$ and 
therefore the CP violating effects by looking at 
the expectation values of the CP-odd observables constructed from the 
momenta of the top quarks and their decay products 
\cite{cpyuan}-\cite{bern}\cite{atwood2}\cite{ext1}-\cite{ext4}. 
In this study, we adopt the following simple observables which 
are constructed from observed momenta and can  
be easily used by experimentists:
\begin{eqnarray} 
A_1&=&E_{l^+}-E_{l^-}~,\\
A_2&=&\vec{p}_{\bar t}\cdot \vec{p}_{l^+}
-\vec{p}_{t}\cdot \vec{p}_{l^-}\equiv O_{\bar t}-Q_{t}~,\\
T_2&=& (\vec{p}_{b}-\vec{p}_{\bar b})\cdot(\vec{p}_{l^+}\times 
\vec{p}_{l^-})~,\\
f_2&=& \frac{\epsilon_{\mu\nu\sigma\rho}
p^\mu_{l^+}p^\nu_{l^-}p^\sigma_{b}p^{\rho}_{\bar b}}
{(p_{l^+}\cdot p_{l^-}p_{b}\cdot p_{\bar b})^{1/2}}~,\\
\hat{O}_L&=&\frac{1}{m_t^3|\vec{P}|^2}
\vec{P}\cdot(\vec{p}_{l^+}\times \vec{p}_{l^-})
\vec{P}\cdot(\vec{p}_{l^+}-\vec{p}_{l^-}) ~, 
\end{eqnarray} 
where all momenta are in the laboratory frame, $E_{l^+}(E_{l^-})$ is the 
energy of $l^+(l^-)(l=l_1=l_2=e,~\mu$, here we do not distinguish $e,~\mu$), 
the subscripts of the momenta  denote the corresponding particles, 
$\vec{P}$ is the momentum of the proton in the case of $p\bar p$ collision. 
$A_1,~A_2,~$ and $T_2$ are studied in Ref.\cite{bern}. $f_2$ and 
$\hat{O}_L$ are used in Refs.\cite{atwood1}\cite{nach2},respectively. 
Because $A_1,~A_2$ are $\hat{T}$-(time reflection\cite{ext3}\cite{timer}) 
even, they are only sensitive 
to the absorptive parts (imaginary parts) of the loop calculations and 
$\hat{d}_t$. On the contrary, $T_2,~f_2,~$ and $\hat{O}_L$ are 
$\hat{T}$-odd and only sensitive to the dispersive parts(real parts).   
$A_1,~T_2,~f_2$ and $\hat{O}_L$ all require the events that two top quarks 
decay semileptonically. $A_2$ uses the events that one top quark  decays 
semileptonically and the other hadronically. Among the above 
observables, only $f_2$ is Lorentz invariant.

While all the above naive observables use only parts of the information,  
the optimal observables studied in Refs.\cite{opt1}-\cite{opt3} use 
the full information in $2Re(M_0^a\delta M^{aA\dagger})$. 
Therefore they are the most effective ones.

In the case of model independent top quark CEDM, 
$2Re(M_0^a\delta M^{aA\dagger})$ contains two terms 
which are proportional to $\hat{d}_t^R$ and $\hat{d}_t^I$, respectively. 
In the MSSM , it depends on the particle masses 
in the loop as well as a multiplicative constant $\lambda_{CP}$. 
However, since our main goal is to search for CP violation induced by 
this $\lambda_{CP}$,  we must assume all the masses in the loop  
are known. In the 2HDM, from the first  term in Eq.(\ref{twohd}) 
which is CP-odd, one can separate 
a factor $\gamma_{CP}=-a\tilde{a}$. Although in the resonant region, 
contributions of the other three terms ( belonging to higher than 
next-to-leading order)  
may be large, they are  still overwhelmed by the tree level and 
the next-to-leading order contributions. As an approximation, we shall 
drop them in the definitions of the optimal observables. 
Therefore we can always separate a constant 
(denote it as $\lambda=\lambda_{CP},~\gamma_{CP},~\hat{d}_t^R, 
~\hat{d}_t^I$)  
from $2Re(M_0^a\delta M^{aA\dagger})$ in all the models considered. 

Apart from some common factors in (\ref{hadroncr}), 
the hadronic cross section can be written as
\begin{eqnarray} 
d\sigma_{p\bar p(p)}&=&\sum\limits_{a=q,g}\int \sum\limits_{color,spin}
\{|M_0^a|^2+\lambda 2Re(M_0^a\delta M^{aA\dagger})\}
\\\nonumber 
& & [f_a^{p}(x_1)f_{\bar a}^{\bar p(p)}(x_2)+
(p\leftrightarrow \bar p(p))]d\Phi dx_1 dx_2~,
\end{eqnarray} 
where $d \Phi$ denotes the phase space. In the following, 
$2Re(M_0^a\delta M^{aA\dagger})$ is calculated by setting 
$\lambda_{CP}=1,~\gamma_{CP}=1,~\hat{d}_t=1$ and $\hat{d}_t=i$ 
in the form factors in each model, respectively. 
The optimal observable is defined as 
\begin{eqnarray} 
O_{opt}&=& \frac
{\sum\limits_{a=q,g}\sum\limits_{color,spin}
 2Re(M_0^a\delta M^{aA\dagger}) 
[f_a^{p}(x_1)f_{\bar a}^{\bar p(p)}(x_2)+
(p\leftrightarrow \bar p(p))]}
{\sum\limits_{a=q,g}\sum\limits_{color,spin}
 |M_0^a|^2 
[f_a^{p}(x_1)f_{\bar a}^{\bar p(p)}(x_2)+
(p\leftrightarrow \bar p(p))]}~.
\end{eqnarray} 
The above observable depends on the parton distribution functions. 
It is inconvenient for practical use. In the $pp$ collision, 
it is not a CP-odd observable due to the asymmetry of quark PDFs
(cf. Appendix C).  Because at the 2 TeV Tevatron, 
the main $t\bar t$ production mechanism is $q\bar q \to t\bar t$, 
and at the 14 TeV LHC, the main process is $ gg\to t\bar t$, we can 
neglect one process at each collider. We consider only $q\bar q\to t\bar t$ 
at the Tevatron and $gg\to t\bar t$ at the LHC in all the following 
calculations. Then we have the optimal observable 
\begin{eqnarray}
\label{opob} 
O_{1}&=& \frac{\sum\limits_{color,spin}
 2Re(M_0^a\delta M^{aA\dagger})}{\sum\limits_{color,spin}|M_0^a|^2 }~.
\end{eqnarray} 
Since the neglected process 
consists of only about $10\%$ of the total cross section, Eq. (\ref{opob}) 
deviates from the truly optimal observable by at most $20\%$. 
$O_{1}$ has the property of Lorentz invariance, one can calculate it 
in any frame. It has the same symmetry property as 
$2Re(M_0^a\delta M^{aA\dagger})$ which is CP-odd. 
In the MSSM and 2HDM, $O_1$ has no definite $\hat{T}$ parity 
and depends on the loop particle masses.  
We note that the optimal observable defined for 
$\hat{d}_t^R$ is only sensitive  to  $\hat{d}_t^R$ independent of 
$\hat{d}_t^I$. The same holds true for $\hat{d}_t^I$. This 
is because the two terms proportional to $d_t^R$ and 
$d_t^I$ have different discrete symmetry $\hat{T}$. 

To calculate $O_1$, we need to know all the momenta of the initial 
and final state particles. That can not always be achieved. 
It is still useful because it can provide us with the upper 
limit signal to noise ratios of any other CP-odd observables. 
Let us first look at the 
case that two top quarks decay semileptonically. The two 
neutrino momenta $p_{\nu_1}$, $p_{\nu_2}$  
will be unknown. But they may be determined  
indirectly by the following eight equations\cite{atwood1}:
\begin{eqnarray}
\label{neutrino}
& &p_{\nu_1}^2=p_{\nu_2}^2=0,~~(p_{\nu_1}+p_{l_1^+})^2=
(p_{\nu_2}+p_{l_2^-})^2=m_W^2,\\ \nonumber
& & (p_{\nu_1}+p_{l_1^+}+p_b)^2=m_t^2,~~
(p_{\nu_2}+p_{l_2^-}+p_{\bar b})^2=m_t^2,~~\\\nonumber
& & (p_{\nu_2}+p_{l_2^-}+p_{\bar b})_{transverse}=
-(p_{\nu_1}+p_{l_1^+}+p_b)_{transverse}~.
\end{eqnarray}
Similar situation exists in the study of $\tau \bar \tau$ production 
in $e^+e^-$ collision\cite{tau}.
The above equations give rise up to fourfold solutions(see Appendix B). 
Considering this ambiguity, we can 
define a modified optimal observable:
\begin{eqnarray}
\label{olob}
O_l&=& \frac{\sum\limits_{color,spin,i}
2Re(M_0^a\delta M^{aA\dagger})\eta_i}
{\sum\limits_{color,spin,i}|M_0^a|^2 \eta_i}~.
\end{eqnarray} 
where the sum $i$ is over all possible solutions of the neutrino 
momenta.$\eta_i=\gamma_i/\hat{s}_i[f_a^p(x_1^i)f_{\bar a}^{p(\bar p)}
(x_2^i)+(p\leftrightarrow p(\bar p))]$  comes from the $t\bar t$ 
phase space, flux factor and PDFs due to different momentum 
reconstruction(cf. Eqs.(\ref{partonc}) and (\ref{hadroncr})). There may be 
possibility that the reconstructed initial parton energy exceeds 
the proton(antiproton) energy. That kind of reconstruction 
should be discarded in the calculations. $O_l$ also depends on 
PDFs. For practical use, we define a non-optimal observable as an 
approximation:  
\begin{eqnarray}
& & O_l'=\sum\limits_{i} O_1~.
\end{eqnarray} 
Again the sum $i$ is over all possible solutions of the neutrino 
momentum\footnote{One can also define $O_l'$ by setting 
$\eta_i=1$ in (\ref{olob}), the numerical difference between 
the two definitions is minor}. 

We now consider that one top quark decays semileptonically and the other 
hadronically. The missing neutrino momentum can be fully reconstructed 
\cite{landsky}.
But because we can not distinguish quark and antiquark jet,
we still have ambiguity of twofold uncertainty. When two quarks 
decay all hadronically, there is  fourfold  uncertainty. We define 
therefore alternatvely the  optimal observables : 
\begin{eqnarray}
O_2&=& \frac{\sum\limits_{color,spin,j}
2Re(M_0^a\delta M^{aA\dagger})}{\sum\limits_{color,spin,j}|M_0^a|^2 }~,\\
O_4&=& \frac{\sum\limits_{color,spin,j'}
2Re(M_0^a\delta M^{aA\dagger})}{\sum\limits_{color,spin,j'}|M_0^a|^2 }~,
\end{eqnarray} 
where the sum $j$ in $O_2$ is over the two possible assignments of the jet 
momenta, and the sum $j'$ in $O_4$ is over the four assignments

The statistical significance of an observable $O$ can be described by 
the signal to noise ratio $r$ defined as 
\begin{eqnarray} 
& & r=<O>/\sqrt{<O^2>}~,
\end{eqnarray} 
where $<O>$,$<O^2>$ are  the expectation value of $O$, $O^2$, respectively:
\begin{eqnarray} 
  & & <O^n>=\frac{\int O^n d\sigma_{p\bar p(p)}}
           {\int d\sigma_{p\bar p(p)}}~. 
\end{eqnarray} 
It is interesting to note that for the optimal observables and unit 
$\lambda$, we always have $<O>=<O^2>$.    
Care must be taken in calculating the $r$ of $A_2$.  Because $A_2$ is 
the difference of two observables ($O_{\bar t}$ and $O_t$)  which 
are calculated using different events(i.e. different independent 
distribution functions), we have $<A_2^2>=<O_{\bar t}^2>+
<O_{t}^2>\approx 2<O_{\bar t}^2>$.  If the experimental error 
comes only from statistics, the number of events 
$N_{event}$ needed to 
observe CP violating effects at $1\sigma$ level (68\% C.L.) satisfys 
$|r|\geq1/\sqrt{N_{event}}$ or $N_{event}\geq 1/r^2$.  To reduce 
the statistical errors, one can combine the measured results of the 
three decay modes:leptonic-leptonic, leptonic-hadronic and 
hadronic-hadronic modes\cite{nedm}. 
Assuming their corresponding number of events are $N_{ll}$,$N_{jl}$ and 
$N_{jj}$, respectively, then we can define a combined signal to noise 
ratio:
\begin{eqnarray}
r_c&=&\sqrt{(N_{ll}r_1^2+N_{jl}r_2^2+N_{jj}r_3^2)/N}, 
\end{eqnarray} 
where $r_1$, $r_2$ and  $r_3$ are the signal to noise ratios of the 
observables which use  
leptonic-leptonic, leptonic-hadronic and hadronic-hadronic events,
respectively. $N=N_{ll}+N_{jl}+N_{jj}$ 
is the total number of events of the three modes. 
Note that $r_c$ depends only on the ratios $N_{ll}/N,~N_{jl}/N$ and 
$N_{jj}/N$, not on $N$.  
The signal is detectable at $1\sigma$ level when $|r_c|\geq1/\sqrt{N}$

\section{ Numerical Results and Conclusions} 
\indent

We first check our calculations with QCD gauge invariance in the 
process $gg\to t\bar t$. That can be 
done by replacing $\epsilon_3,~\epsilon_4$ with $p_3,~p_4$. 
We find that the correction amplitude is consistent with gauge 
invariance.  
Then we check our calculations with known results 
in the literature. Our parton level results are all in agreement with 
Refs.\cite{schmidt}\cite{bern}. By using $\sqrt{s}=40$ TeV, 
$m_t=160$ GeV and 
$10^7$ sample of leptonic events, we find the $O_1$ limit on $d_t^R$ 
is $2.8\times 10^{-20}~cm~g_s$ which is very close to $\lambda_{min}$ 
in Ref.\cite{atwood1}. That means that taking 
the top quark spin in its rest frame to be  
in the direction of lepton momentum is feasible. We can also reproduce 
the results of $\hat{O}_L$ in Ref.\cite{nach2} with $\mu_t'=0$
 by including both  
$q\bar q\to t\bar t$ and $gg\to t \bar t$ processes in the calculations. 

As mentioned previously, in the following calculations, we 
consider only $q\bar q\to t\bar t$ at the Tevatron and $gg\to t \bar t$ at 
the LHC. The parton distribution functions of MRS set A'\cite{MRSA}
with scale $Q^2=m_t^2$   
are used\footnote{It should be noted that all the results are insensitive 
to the PDF and $Q^2$ choices. CTEQ4M\cite{cteq4} gives similar results.}. 
$m_t$ is take to be $176$ GeV.   
To look at the largest possible effects, we set the CP-violating 
parameters $\lambda_{CP}$ and $\gamma_{CP}$ to be of order 1, namely, 
$\lambda_{CP}=1$(this needs $\theta_t=\pi/4$, cf. Appendix A), 
$\gamma_{CP}=1$. In the MSSM and 2HDM, we treat 
the SUSY particle masses and Higgs boson mass as free parameters 
allowed by current experiments. We assume that all the squarks 
except for the light stop $\tilde{t}_1$ to be degenerate.  
The light stop mass is required to be  above $50$ GeV\cite{lstop}.
We choose gluino mass $m_{\tilde{g}}$ and Higgs boson mass 
$\geq 100$ GeV.  
 
The signal to noise ratio results at the 2 TeV $p\bar p$ Tevatron and 
at the 14 TeV $pp$ LHC are summarized in Table I.--Table V.
We do not present the results for 2HDM at the Tevatron 
because the effects are too small unless the Higgs boson mass is less than 
$100$ GeV. We denote the signal to noise ratio $r$ of an observable $O$ as 
$r(O)$. All the tables show that $r(O_l')$ is slightly smaller 
than $r(O_1)$ and $r(O_2)$ is  about $3/4$ 
of  $r(O_1)$, $r(O_4)$ about $1/2$ of $r(O_1)$. 
Therefore, $O_l'$ is a good approximation for $O_1$. 

Table I. shows the results at the Tevatron from five sets of 
SUSY parameters.  
We see that the naive observables  
have signal to noise ratios all $\le 1\%$. It is difficult to observe 
such small effects at the Tevatron. The optimal observables, 
on the other hand,  
have $r\geq 1\%$ as long as the gluino mass is around $200$ GeV
\footnote{The value is close to $m_t$, so that the gluino threshold 
is close to the top quark one. The  CP violating effects 
are large due to the threshold effects.}. 
They are about $3-10$ times more
effective than the naive ones.  The combined signal to noise 
ratios of $O_l'$,$O_2$ and $O_4$ are all between these of $O_2$ and 
$O_4$. Because $N_{ll}$ is small compared with $N_{jl}$ and 
$N_{jj}$, the combined $r$ is just a weighted average of 
these of $O_2$ and $O_4$. 
Assuming we can obtain $30~fb^{-1}$ integrated luminosity,  
then  the total number of fully reconstructed $t\bar t$ events 
is about $3.5\times 10^4$\cite{tev200}.  
It is reasonable to assume that we can have purely hadronic 
events $N_{jj}=2\times 10^4$, hadronic-leptonic events $N_{jl}=1.2
\times 10^4$ and purely leptonic events $N_{ll}=0.2\times 10^4$. 
The corresponding 1 $\sigma$ level statistical 
errors are $r_{jj}=0.71\times 10^{-2},
~ r_{jl}=0.91\times 10^{-2},~ r_{ll}=2.2\times 10^{-2}$.    
The combined error is $0.54\%$. In the Table, $O_l'$ has $r>r_{ll}$ only 
when $m_{\tilde{t}_1}$ is around 50 GeV and 
$m_{\tilde{g}}$ is around 200 GeV. We always have $r(O_2)>r_{jl}$,  
$r(O_4)>r_{jj}$ when $m_{\tilde{g}}\sim200$ GeV. Therefore, 
it is possible to detect the CP violating effects 
by using optimal observables $O_2$ and $O_4$ when 
$m_{\tilde{g}}$ is around 200 GeV.  
The effects are detectable by combined measurement  
when $m_{\tilde{g}}$ is in the range 100-300 GeV.

Table II. is the results at the LHC in MSSM. 
It shows the same features as Table I. However, at the LHC, 
the numbers of events are much larger than those at the Tevatron. 
With $150~fb^{-1}$ integrated luminosity, we can assume 
$N_{jj}=10^7$, $N_{jl}=6\times 10^6$ and $N_{ll}=10^6$. We further 
assume the experimental systematic errors are below the statistical  
ones. In addition, there are also theoretical uncertainties coming from  
possible non-CP violating contaminations at the $pp$ LHC. 
Because the initial $pp$ state is not a CP eigenstate, CP 
conserving interactions can produce CP asymmetry 
effects in $t\bar t$ final state. 
We present a general analysis of the contaminations to the 
Lorentz invariant observables in Appendix C.  We 
find that within the framework of parton model, 
there are no contaminations to these observables. 
$A_1,~A_2$ and $T_2$ are discussed in Refs. \cite{peskin}\cite{bern}.    
They are well below the signals. 
Therefore, the 1 $\sigma$ level errors are $r_{jj}=0.32\times 10^{-3},
~r_{jl}=0.41\times 10^{-3},~ r_{ll}=1.0\times 10^{-3}$.    
The combined error is $0.24\times 10^{-3}$. 
The naive observable $T_2$ is on the margin to be detectable. $A_1$ and 
$f_2$ are better than $T_2$. They can be used to observe 
$\lambda_{CP}$ at few $\times 10^{-1}$ when $m_{\tilde{g}}$ is 
within the range $100-400$ GeV.  The  observable $O_l'$ is 2-5 times better. 
All $r(O_2)\geq 10~r_{jl}$ and $r(O_4)\geq 10~r_{jj}$ for 
$m_{\tilde{g}}\sim 100-300$ GeV. 
We can pin down $\lambda_{CP}$ to $10^{-1}$ by using  these 
optimal observables.

In Table III., we give the results of signal to noise ratios 
in 2HDM at the LHC. It is obvious that $A_1$ and $T_2$ are only 
detectable and can not be used to put limit on $\gamma_{CP}$. 
$f_2$ is 2 times better which may be used to limit $\gamma_{CP}$ to 
$(3-4)\times 10^{-1}$. All the optimal observables have 
$r$ large than about 10 times statistical errors. Therefore, 
they will put a limit of order $10^{-1}$ on $\gamma_{CP}$. 

Table IV. and V. are the results of model-independent 
top quark CEDM. An overall feature of the two tables is that both at 
Tevatron and LHC, the accuracies of $d_t^I$ are better 
than those of $d_t^R$. With the above assumed numbers of events  
at the Tevatron, the best 
limits on $d_t^R$ and $d_t^I$ are $2.4\times 10^{-18}cm~g_s$ and  
$1.1\times 10^{-18}cm~g_s$,respectively,  by using $O_2$.    
Also by using $O_2$ at the LHC, we can obtain  the 
limits of $5.2\times 10^{-20}cm~g_s$ and $2.5\times 10^{-20}cm~g_s$ on 
$d_t^R$ and $d_t^I$, respectively.    

To summarize, we have studied 
CP violating effects in top quark pair production at the future 
2 TeV $p\bar p$ Tevatron and 14 TeV $pp$ LHC colliders. 
Three kinds of  CP violating sources:the SUSY CP-odd phase of the stop 
trilinear soft breaking term, $arg(A_t)$, the CP-odd parameter in 
2HDM, and the model-independent top quark CEDM are investigated. 
Optimal observables as well as simple observables are used. 
The optimal obervables are usually $2-10$ times more effective 
than the naive ones. It is possible to observe CP violating
effects from $arg(A_t)$ in top quark pair production at the 2 TeV Tevatron 
with $\sim 30 fb^{-1}$ integrated luminosity when $m_{\tilde{g}}
\sim 200$ GeV. If combined measurement is applied, the range 
of gluino mass in which the CP violating effects are detectable 
is $\sim 100-300$ GeV. The LHC with $150fb^{-1}$ can put a limit of 
order $10^{-1}$ on $\lambda_{CP}$ (therefore on $arg(A_t)$) 
in MSSM and $\gamma_{CP}$  in 2HDM by using optimal observables
provided the experimental errors are sufficient small.  
The CEDM of the top quark can be measured to an accuracy of 
$10^{-18} ~cm~g_s$ at the Tevatron and few $\times 10^{-20} ~cm~g_s$ 
at the LHC. More accurate measurement on $d_t^I$ can be obtained 
than on $d_t^R$ with given number of events.

\begin{table}
\caption{} 
Signal to noise ratio $r$  
in $p \bar p \to t \bar t +X $ at the 2 TeV Tevatron in the MSSM 
with $\lambda_{CP}=1$,  
for five sets of SUSY parameters labeled by 
$(m_{\tilde{t}_1},m_{\tilde{t}_2}=m_{\tilde{q}},m_{\tilde{g}})$ GeV. 
The combined results are for $O_l'$, $O_2$ and $O_4$. 
\small{
\begin{center}
\begin{tabular}{|c|c|c|c|c|c|c|c|c|c|}\hline
&$A_1$ & $A_2$ &$T_2$& $f_2$ & $O_1$ &$O_l'$ & $O_2$ &  $O_4$     
& combined  \\ \hline
(100,500,100) & $0.25$\% & $-0.23$\% & $-0.16$\% & $-0.16$\% & 0.82\%   
& 0.79\% & 0.62\% & 0.45\% &{\bf 0.54\%}\\\hline
(100,500,200) & $0.59$\% & $-0.42$\% & $0.08$\% & $0.12$\% & 1.82\%  
 & 1.73\% & 1.41\%   & 1.04\% & {\bf 1.18\%} \\\hline
(100,500,300) & $0.12$\%& $-0.17$\%& $0.13$\%& $0.11$\% & 0.78\%
& 0.74\%& 0.56\%& 0.40\%&{\bf 0.49\%} \\\hline
(50,500,200) & $0.80$\%& $-0.58$\%& $0.08$\% & $0.14$\% 
& 2.49\%& 2.37\%& 1.93\%& 1.43\% &{\bf 1.69\%}   \\\hline
(100,1000,200)& $0.58$\% & $-0.42$\% & $0.11$\% & $0.15$\% 
& 1.77\%  & 1.68\% & 1.36\%  & 1.01\% &{\bf 1.19\%}  
\\\hline
\end{tabular}
\end{center}
}
\end{table}
\begin{table}
\caption{ }
Signal to noise ratio $r$  
in $p p \to t \bar t +X $ at the LHC in the MSSM 
with $\lambda_{CP}=1$,  
for six sets of SUSY parameters labeled by 
$(m_{\tilde{t}_1},m_{\tilde{t}_2},m_{\tilde{g}})$ GeV. 
The combined results are for $O_l'$, $O_2$ and $O_4$. 
\small{
\begin{center}
\begin{tabular}{|c|c|c|c|c|c|c|c|c|}
\hline
&$A_1$ &$T_2$ & $f_2$ & $O_1$ &$O_l'$& $O_2$& $O_4$& combined  \\\hline
(100,500,100) & 
$-0.12$\% &0.11\% & 0.22\% & 0.92\% &0.81\% &0.71\% & 0.53\% &{\bf 0.62\%}  
\\\hline
(100,500,200) & 
  0.41\% &$-0.13$\% &$-0.35$\% & 1.82\% &1.61\% &1.41\% 
& 1.05\% &{\bf 1.23\%}\\\hline
(100,500,300) & 
0.17\% &$-0.13$\% &$-0.29$\%&0.85\%&0.72\%&0.62\%& 0.44\%&{\bf 0.53\%}
\\\hline
(100,500,400) & 
0.06\%& $-0.08\%$& $-0.17\%$ &0.41\% & 0.34\% & 0.29\% & 0.20\% & {\bf 0.24\%}
\\\hline
(50,500,200) & 
0.63\%  &$-0.14$\% &  $-0.42$\% & 2.50\%  & 2.22\%  & 1.97\%  &1.47\%  &
{\bf 1.71\%}\\\hline
(100,1000,200) & 
  0.47\% &$-0.17$\% &
 $-0.48$\%& 2.12\% & 1.88\% &  1.62\% &  1.20\% &{\bf 1.41\%}
\\\hline
\end{tabular}
\end{center}
}
\end{table}

\begin{table}
\caption{ }
Signal to noise ratio $r$  
in $p p \to t \bar t +X $ at the LHC in the 2HDM 
with $\gamma_{CP}=1$.   
The combined results are for $O_l'$, $O_2$ and $O_4$. 
\small{
\begin{center}
\begin{tabular}{|c|c|c|c|c|c|c|c|c|}
\hline
$m_\phi$(GeV)& 
$A_1$ & $T_2$ & $f_2$ & $O_1$ & $O_l'$&$O_2$ & $O_4$ & combined \\\hline
100 &$-0.20$\% & 0.10\% & 0.28\% & 1.09\%  & 0.97\%  & 0.82\%  
& 0.60\%  &{\bf 0.71\%}\\\hline
200 &$-0.16$\% & 
0.12\% & 0.30\% &  0.96\%  &0.85\%  &0.69\%  & 0.49\%  & {\bf 0.59\%}
\\\hline
300 & $-0.15$\% &0.14\% & 
0.36\% &  1.12\%  & 0.97\%  &0.78\%  &0.54\%  &{\bf 0.66\%}\\\hline
400 &  $-0.13$\% & 0.20\% & 0.45\% &  1.55\%  & 
 1.52\%  &  1.31\%  &  0.92\%  & {\bf 1.11\%}\\\hline
500 &$-0.05$\% & 0.15\% &
0.31\% &0.91\%  &0.87\%  &0.76\% &0.53\% &{\bf 0.64\%}\\\hline
\end{tabular}
\end{center}
}

\end{table}

\begin{table}
\caption{ }
Signal to noise ratio $r$  
at the Tevatron and LHC with model independent top quark 
CEDM $d_t^R=1~$GeV$^{-1}~g_s=1.97\times10^{-14}cm~g_s$, and  
the accuracies with which   $d_t^{R}$  can be measured 
at the Tevatron and LHC with assumed numbers of events given 
in the text.$d_t^{R}$ is given in unit of $10^{-18}~cm~g_s$. 
$O_1$ is given only for leptonic events. The combined results 
are for $O_l'$, $O_2$, $O_4$. 
\small{
\begin{center}
\begin{tabular}{|c|c|c|c|c|c|c|c|c|c|}
\hline
\multicolumn{2}{|c|}{~} 
&$T_2$ & $f_2$ & $\hat{O}_L$& $O_1$ & $O_l'$ 
& $O_2$ & $O_4$ & combined
\\ \hline
Tevatron & $r$ & 39.9 & 40.9 & $-35.2$ & 124 & 88 & 75 & 47 &{\bf 61}
\\\cline{2-10}
& $d_t^R$ & 10.9 & 10.6 & $12.3$ & $3.5$& 4.9  & 2.4 & 3.0 & {\bf 1.7}
\\ \hline
LHC & $r$ & $-53.6$ & $-112$ &  & 238 & 199  & 154 & 99 & {\bf 128}
\\ \cline{2-10}
& $d_t^R$ & $0.37 $ & $0.18$ &  & $0.08$& 0.10  & 0.052 & 0.064 &{\bf 0.037} 
\\ \hline
\end{tabular}
\end{center}
}
\end{table}

\begin{table}
\caption{ }
The same as Table IV., but for $d_t^I$. 
\small{
\begin{center}
\begin{tabular}{|c|c|c|c|c|c|c|c|c|}
\hline
\multicolumn{2}{|c|}{~}    
&$A_1$ & $A_2$ & $O_1$ & $O_l'$ 
& $O_2$ & $O_4$ & combined
\\\cline{1-9}
Tevatron     & $r$ & $-80.5$ & 57.4 & $214$ & 179 & 169 & 127  & 
{\bf 146}\\\cline{2-9}
& $d_t^I$ & 5.4 & $4.4^b$ & $2.0$ & 2.4  & 1.1 & 1.1 & {\bf 0.73}
\\ \hline
LHC & $r$ & $83.2$ & $-21.1$ & 381 & 332  & 332 & 248 &{\bf 282}
\\ \cline{2-9}
& $d_t^I$ & $0.24 $ & $5.8^b$ & $0.052$ & $0.059$ & 0.025  & 0.025 & 
{\bf 0.017}\\ \hline
\end{tabular}
\end{center}
}
$^b$ Note that for $A_2$, only half of $N_{jl}$ can be used.  
\end{table}

\vspace{2.cm}
\begin{center}
{\bf Acknowledgements} 
\end{center}

The author is grateful to  O. Nachtmann and A.Brandenburg for 
helpful suggestions and discussions and to O.Nachtmann for carefully 
reading the manuscript. 
The author also benefits from communications with W.Hollik. 
This work is financially supported by the AvH Foundation of Germany.

\eject
\vspace{2.cm}

\appendix
\section{Form factors}

We give here the non-zero form factors for the matrix elements  
appeared in the text.  
They are written in terms of the conventional one-, two-, three- and 
four-point scalar loop integrals defined in Ref.\cite{veltman}.  

In the following, the form factors are given in the MSSM.
\begin{eqnarray}
D^s &= & \sum\limits_{j=1,2}
\{F_1C_{11}
(-p_2,k,m_{\tilde{t}_j},m_{\tilde{g}},m_{\tilde{g}})\\\nonumber
& &+F_2(C_0+C_{11})
(-p_2,k,m_{\tilde{g}},m_{\tilde{t}_j},m_{\tilde{t}_j})
\}\frac{\alpha_s}{4\pi}m_{\tilde{g}}(-1)^{j+1}i\lambda_{CP}
\end{eqnarray} 
\begin{eqnarray}
f^{DB} &= & \sum\limits_{j=1,2}\{D_{13}
(-p_2,p_4,p_3,m_{\tilde{t}_j},m_{\tilde{g}},m_{\tilde{q}},m_{\tilde{g}})
\}\frac{\alpha_s}{4\pi}m_{\tilde{g}}(-1)^{j+1}i\lambda_{CP}
\end{eqnarray}

\begin{eqnarray}
f^{s1}_1&=&-2D^s(p_1\cdot p_4-p_2\cdot p_4)/\hat{s}~,\\
f^{s1}_8&=& 4D^s/\hat{s}~,\\
f^{s1}_9&=& -4D^s/\hat{s}~.
\end{eqnarray}

\begin{eqnarray}
f^{s2}_1
&=&\sum\limits_{j=1,2}\{ C_0(-p_2,k,m_{\tilde{g}},m_{\tilde{t}_j},
m_{\tilde{t}_j})\}\frac{\alpha_s}{4\pi}m_{\tilde{g}}(-1)^{j+1}i\lambda_{CP}
\end{eqnarray}

\begin{eqnarray}
f^{self,t}_2
&=&\sum\limits_{j=1,2}\{ -B_0(\hat{t},m_{\tilde{g}}^2,m_{\tilde{t}_j}^2)
+B_0(m_t^2,m_{\tilde{g}}^2,m_{\tilde{t}_j}^2)\}\\ \nonumber
& &\times \frac{(F_1+F_2)\alpha_s}{4\pi}m_{\tilde{g}}
(-1)^{j+1}i\lambda_{CP}/(\hat{t}-m_t^2).
\end{eqnarray}

\begin{eqnarray}
f^{v1,t}_2
&=&\sum\limits_{j=1,2}\{ -C_0(-p_2,p_4,m_{\tilde{t}_j},
m_{\tilde{g}},m_{\tilde{g}})
\}\frac{F_1\alpha_s}{4\pi}m_{\tilde{g}}(-1)^{j+1}i\lambda_{CP},\\\nonumber
f^{v1,t}_{16}&=& 2 D^{v}/(\hat{t}-m_t^2).
\end{eqnarray}

\begin{eqnarray}
f^{v2,t}_2&=&\sum\limits_{j=1,2}\{ -C_0(p_1,-p_3,m_{\tilde{t}_j},
m_{\tilde{g}},m_{\tilde{g}})
\}\frac{F_1\alpha_s}{4\pi}m_{\tilde{g}}(-1)^{j+1}i\lambda_{CP},\\
f^{v2,t}_{9}&=& 4 D^{v}/(\hat{t}-m_t^2),\\
f^{v2,t}_{13}&=& 2 D^{v}/(\hat{t}-m_t^2),
\end{eqnarray}
where 
\begin{eqnarray}
D^v &= & \sum\limits_{j=1,2}
\{F_1C_{11}
(-p_2,p_4,m_{\tilde{t}_j},m_{\tilde{g}},m_{\tilde{g}})\\ \nonumber
& & +F_2(C_0+C_{11})(-p_2,p_4,m_{\tilde{g}},m_{\tilde{t}_j},
m_{\tilde{t}_j})
\}\frac{\alpha_s}{4\pi}m_{\tilde{g}}(-1)^{j+1}i\lambda_{CP}
\end{eqnarray}

\begin{eqnarray}
f_1^{box1,t}&=&\sum\limits_{j=1,2}\{-4D_{27}\}
\frac{\alpha_s}{4\pi}m_{\tilde{g}}(-1)^{j+1}i\lambda_{CP}F_1\\
f_2^{box1,t}&=&\sum\limits_{j=1,2}\{ m_t^2(D_0+2D_{12}+2D_{13}+D_{24}+D_{25})
-m_{\tilde{g}}^2D_0-2p_2\cdot p_4(2D_{12}\\ \nonumber
& & +D_{24}-D_{26})
-2p_1\cdot p_4(D_{25}-D_{26})+4D_{27}\}
\frac{\alpha_s}{4\pi}m_{\tilde{g}}(-1)^{j+1}i\lambda_{CP}F_1\\
f_7^{box1,t}&=&\sum\limits_{j=1,2}\{-4D_{26}\}
\frac{\alpha_s}{4\pi}m_{\tilde{g}}(-1)^{j+1}i\lambda_{CP}F_1\\
f_8^{box1,t}&=&\sum\limits_{j=1,2}\{4(D_{25}-D_{26})\}
\frac{\alpha_s}{4\pi}m_{\tilde{g}}(-1)^{j+1}i\lambda_{CP}F_1\\
f_9^{box1,t}&=&\sum\limits_{j=1,2}\{4(D_{12}+D_{24}-D_{26})\}
\frac{\alpha_s}{4\pi}m_{\tilde{g}}(-1)^{j+1}i\lambda_{CP}F_1\\
f_{10}^{box1,t}&=&\sum\limits_{j=1,2}\{-4(D_{13}+D_{26})\}
\frac{\alpha_s}{4\pi}m_{\tilde{g}}(-1)^{j+1}i\lambda_{CP}F_1\\
f_{13}^{box1,t}&=& \sum\limits_{j=1,2}\{2D_{12}\}
\frac{\alpha_s}{4\pi}m_{\tilde{g}}(-1)^{j+1}i\lambda_{CP}F_1\\
f_{14}^{box1,t}&=&\sum\limits_{j=1,2}\{ -2D_{13}\}
\frac{\alpha_s}{4\pi}m_{\tilde{g}}(-1)^{j+1}i\lambda_{CP}F_1\\
f_{15}^{box1,t}&=& \sum\limits_{j=1,2}\{-2D_{13}\}
\frac{\alpha_s}{4\pi}m_{\tilde{g}}(-1)^{j+1}i\lambda_{CP}F_1\\
f_{16}^{box1,t}&=& \sum\limits_{j=1,2}\{2D_{12}\}
\frac{\alpha_s}{4\pi}m_{\tilde{g}}(-1)^{j+1}i\lambda_{CP}F_1
\end{eqnarray}
The above $D$ functions have the arguments 
$(-p_2,p_4,p_3,m_{\tilde{t}_j},m_{\tilde{g}},m_{\tilde{g}},
m_{\tilde{g}}) $.

\begin{eqnarray}
f_1^{box2,t}&=&\sum\limits_{j=1,2}\{-4D_{27}\}
\frac{\alpha_s}{4\pi}m_{\tilde{g}}(-1)^{j+1}i\lambda_{CP}F_2\\
f_7^{box2,t}&=&\sum\limits_{j=1,2}\{-4(D_{13}+D_{26})\}
\frac{\alpha_s}{4\pi}m_{\tilde{g}}(-1)^{j+1}i\lambda_{CP}F_2\\
f_8^{box2,t}&=&\sum\limits_{j=1,2}\{4(D_{25}-D_{26})\}
\frac{\alpha_s}{4\pi}m_{\tilde{g}}(-1)^{j+1}i\lambda_{CP}F_2\\
f_9^{box2,t}&=&\sum\limits_{j=1,2}\{4(D_0+2D_{12}+D_{24}-D_{26})\}
\frac{\alpha_s}{4\pi}m_{\tilde{g}}(-1)^{j+1}i\lambda_{CP}F_2\\
f_{10}^{box2,t}&=&\sum\limits_{j=1,2}\{-4(D_{13}+D_{26})\}
\frac{\alpha_s}{4\pi}m_{\tilde{g}}(-1)^{j+1}i\lambda_{CP}F_2
\end{eqnarray}
The above  $D$ functions have the arguments 
$(-p_2,p_4,p_3,m_{\tilde{g}},m_{\tilde{t}_j},m_{\tilde{t}_j},
m_{\tilde{t}_j}) $.

\begin{eqnarray}
f_1^{box3,t}&=&\sum\limits_{j=1,2}\{-4D_{27}\}
\frac{\alpha_s}{4\pi}m_{\tilde{g}}(-1)^{j+1}i\lambda_{CP}F_1F_2\\
f_7^{box3,t}&=&\sum\limits_{j=1,2}\{4(-D_{23}+D_{26})\}
\frac{\alpha_s}{4\pi}m_{\tilde{g}}(-1)^{j+1}i\lambda_{CP}F_1F_2\\
f_8^{box3,t}&=&\sum\limits_{j=1,2}\{4(-D_{25}+D_{26})\}
\frac{\alpha_s}{4\pi}m_{\tilde{g}}(-1)^{j+1}i\lambda_{CP}F_1F_2\\
f_9^{box3,t}&=&\sum\limits_{j=1,2}\{4(D_{12}-D_{13}+D_{24}-D_{25})\}
\frac{\alpha_s}{4\pi}m_{\tilde{g}}(-1)^{j+1}i\lambda_{CP}F_1F_2\\
f_{10}^{box3,t}&=&\sum\limits_{j=1,2}\{-4(D_{11}-D_{12}+D_{21}-D_{24})\}
\frac{\alpha_s}{4\pi}m_{\tilde{g}}(-1)^{j+1}i\lambda_{CP}F_1F_2\\
f_{13}^{box3,t}&=&\sum\limits_{j=1,2}\{ 2(D_{12}-D_{13})\}
\frac{\alpha_s}{4\pi}m_{\tilde{g}}(-1)^{j+1}i\lambda_{CP}F_1F_2\\
f_{15}^{box3,t}&=&\sum\limits_{j=1,2}\{ -2(D_{11}-D_{12})\}
\frac{\alpha_s}{4\pi}m_{\tilde{g}}(-1)^{j+1}i\lambda_{CP}F_1F_2
\end{eqnarray}
The above $D$ functions have the arguments 
$(-p_2,p_4,-p_1,m_{\tilde{t}_j},m_{\tilde{g}},m_{\tilde{g}},
m_{\tilde{t}_j}) $.

In the above, the color factors
$\displaystyle F_1=\frac{3}{2}~,~~F_2=-\frac{1}{6}$, and 
$\lambda_{CP}=2Im(a_1b_1^\star)=\sin2\theta_t\sin\beta_t$ .

The form factors of 2HDM can be obtained from those of 
MSSM by setting $j=1~,~~F_1= 1~, ~~F_2= 0~,~~f^{DB}
=f_n^{s2}=0~$, 
and the following substitutions:
\begin{eqnarray}
& & \lambda_{CP}\to 2a\tilde{a}=-2\gamma_{CP}~,\\
& & m_{\tilde{g}}\to m_t~,~~m_{\tilde{t}_1}\to m_{\phi_1}~,\\
& & \frac{\alpha_s}{4\pi}\to \frac{\sqrt{2}m_t^2 G_F}{16\pi^2}~,
\end{eqnarray}

In 2HDM, the form factors of Fig.2 (n) defined in Eq.(\ref{res}) are
\begin{eqnarray}
f^{sr}_1&=& \frac{\sqrt{2}m_t^2 G_F}{16\pi^2}\{4m_t[m_t^2C_0
-p_3\cdot p_4(2C_{22}-2C_{23}+C_0)+\frac{1}{2}]\}~,\\
f^{sr}_2&=& \frac{\sqrt{2}m_t^2 G_F}{16\pi^2}\{4m_t[C_0
+4(C_{22}-C_{23})]\}~,\\
f^{sr}_3&=& \frac{\sqrt{2}m_t^2 G_F}{16\pi^2}\{4m_tC_0\}~,
\end{eqnarray}
where the C function arguments 
are $(p_4,-k,m_t,m_t,m_t)$.

The constant model-independent form factors are:

\begin{eqnarray}
D^s&=&i\hat{d}_t~,\\
f^{s}_1&=&-2i\hat{d}_t(p_1\cdot p_4-p_2\cdot p_4)/\hat{s}~,\\
f^{s}_8&=& 4i\hat{d}_t/\hat{s}~,\\
f^{s}_9&=& -4i\hat{d}_t/\hat{s}~,\\
f^{v,t}_2&=& 2i\hat{d}_t~,\\
f^{v,t}_9&=& 4i\hat{d}_t/(\hat{t}-m_t^2)~,\\
f^{v,t}_{13}&=& 2i\hat{d}_t/(\hat{t}-m_t^2)~,\\
f^{v,t}_{16}&=& 2i\hat{d}_t/(\hat{t}-m_t^2)~,\\
f^{b}_1&=& -i\hat{d}_t~,\\
f^{b}_2&=& i\hat{d}_t~,
\end{eqnarray}
where $\hat{d}_t$ is defined in the text.

Other definitions:
\begin{eqnarray}
& & k=p_1+p_2=p_3+p_4,\;\;\hat{s}=k^2,\;\;
\hat{t}=q^2=(p_2-p_4)^2,\;\;\hat{u}=(p_2-p_3)^2,\\ \nonumber
& & \Gamma^\mu=(-p_4+p_3)^\mu\epsilon_3\cdot\epsilon_4+(2p_4+p_3)
\cdot\epsilon_3\epsilon_4^\mu-(2p_3+p_4)\cdot\epsilon_4\epsilon_3^\mu \;.
\end{eqnarray}

\section{Solution of the neutrino momenta}

We give here briefly the method of solving the neutrino 
momenta in Eqs.(\ref{neutrino}). Let $p_{l_1^+}=p_l,~
p_{l_2^-}=p_{l}'$, $p_{\bar b}=p_b'$, 
$p_{\nu_1}=(E,X,Y,Z),~
p_{\nu_2}=(E',X',Y',Z'),~p_{lb}=p_{l}+p_b$, $ p_{lb}'=
p_{l}'+p_{b}'$, $ p_0=-(p_b+p_b'+p_l+p_l')$. 
Then we can get
\begin{eqnarray}
& & X= a E+b Z+\delta ~,~~Y= c E+d Z+\xi~, \\\nonumber
& & X'= a' E'+b' Z'+\delta' ~,~~Y'= c' E'+d' Z'+\xi'~,
\end{eqnarray}
where 
\begin{eqnarray}
& & a=\frac{E_{lb}p_l^y-E_lp_{lb}^y}{\Delta_1}~,~~
b=\frac{-p_{lb}^zp_l^y+p_l^zp_{lb}^y}{\Delta_1}~,~~\\ \nonumber
& & c=\frac{-E_{lb}p_l^x+E_lp_{lb}^x}{\Delta_1}~,~~
d=\frac{p_{lb}^zp_l^x-p_l^zp_{lb}^x}{\Delta_1}~,~~\\ \nonumber
& & \delta=\frac{-(m_t^2-p_{lb}^2)/2p_l^y+m_W^2/2p_{lb}^y}{\Delta_1}~,~~
\xi=\frac{(m_t^2-p_{lb}^2)/2p_l^x-m_W^2/2p_{lb}^x}{\Delta_1}~,~~\\\nonumber
& & \Delta_1=p_{lb}^xp_l^y-p_{lb}^yp_l^x ~, 
\end{eqnarray} 
and $a',b',c',d',\delta',\xi'$ are obtained from the above equations by 
the substitutions of momenta without $'$ to those with $'$.
We further express $E, E'$ in terms of $Z,Z'$:
\begin{eqnarray}
& & E=fZ+gZ'+h~,~~E'=f'Z+g'Z'+h'~,~~
\end{eqnarray}
where 
\begin{eqnarray}
& & f=\frac{-bc'+da'}{\Delta_2},~~g=\frac{-b'c'+d'a'}{\Delta_2},~~
h=\frac{(p_0^x-\delta-\delta')c'+a'(p_0^y-\xi-\xi')}{\Delta_2},~~\\ \nonumber
& & f'=\frac{-ad+cb}{\Delta_2},~~g'=\frac{-ad'+cb'}{\Delta_2},~~
h'=\frac{(p_0^y-\xi-\xi')a+c(p_0^x-\delta-\delta')}{\Delta_2},~~\\ \nonumber
& & \Delta_2=ac'-ca'~. 
\end{eqnarray}
Inserting the above expressions into $E^2=X^2+Y^2+Z^2$, and $
E'^2=X'^2+Y'^2+Z'^2$, we get the following two quadratic equations 
for $Z,~Z'$:
\begin{eqnarray}
\label{solv1}
& &AZ^2+BZZ'+CZ'^2+D+UZ+VZ'=0~,~~\\ 
\label{solv2}
& &A'Z'^2+B'ZZ'+C'Z^2+D'+U'Z'+V'Z=0~,~~
\end{eqnarray}
with 
\begin{eqnarray}
& & A=f^2-1-(af+b)^2-(cf+d)^2~,
~~B=2[fg-ag(af+b)-cg(cf+d)]~,~~\\\nonumber
& & C=g^2-a^2g^2-c^2g^2~,~~
D =h^2-(ah+\delta)^2-(ch+\xi)^2~,~~\\\nonumber
& & U=2[fh-(ah+\delta)(af+b)-(ch+\xi)(cf+d)]~,~~\\\nonumber
& & V=2[gh-ag(ah+\delta)-cg(ch+\xi)]~,~~
\end{eqnarray}
and $A',B',C',D',U',V'$ can be gotten from the above equations 
by replacing the variables without $'$ with the ones with $'$ and 
interchanging $f',~g'$.

Eqs.(\ref{solv1})(\ref{solv2}) can obviously give at most four real 
solutions for $(Z,Z')$. They can be solved by standard methods: 
solving $Z'$ in (\ref{solv1}) as functions of $Z$, we get  
two expressions for $Z'$, then inserting them separately 
into (\ref{solv2}), we shall get the same quartic equation of $Z$ 
which has up to four real solutions. Although those solutions 
will be doubled when we calculate $Z'$ with the two expressions 
obtained from (\ref{solv1}), half of them are false solutions  
to (\ref{solv2}).

 
\section{Analysis on the contaminations to CP violating 
effects in $pp$ collision }

We assume the partons inside the proton have no intrinsic transverse 
momenta and denote the parton level reaction as 
$a(x_1)\bar a(x_2)\to t\bar t$ 
with $x_1$ and $x_2$ being the Bjorken scale parameters.
The hadronic level cross section can be written as
\begin{eqnarray}
d\sigma&=&2 f_{a}^p(x_1)f_{\bar a}^p(x_2) d\hat{\sigma}(a\bar a\to t\bar t)
dx_1dx_2/(1+\delta_{a,\bar a})\\\nonumber
&=&\{[f_{a}^p(x_1)f_{\bar a}^p(x_2)+f_{a}^p(x_2)f_{\bar a}^p(x_1)]  
+[f_{a}^p(x_1)f_{\bar a}^p(x_2)-f_{a}^p(x_2)f_{\bar a}^p(x_1)]\}\\\nonumber
& & \times d\hat{\sigma}(a\bar a\to t\bar t)/(1+\delta_{a,\bar a})
dx_1dx_2,
\end{eqnarray}
where $f^{p}_{a},~f^{p}_{\bar a}$  are 
the parton distribution functions of $a$ and $\bar a$ in proton.
The function $F^+(x_1,x_2)
=f_{a}^p(x_1)f_{\bar a}^p(x_2)+f_{a}^{p}(x_2)f_{\bar a}^{p}(x_1)$
is CP-even(its CP transformation is just the interchanging of $x_1,~x_2$), 
so that no contamination comes from it. Now 
we look at the function $F^-(x_1,x_2)
=f_{a}^p(x_1)f_{\bar a}^p(x_2)-f_{a}^{p}(x_2)f_{\bar a}^{p}(x_1)$. 
Contaminations should come from $F^-(x_1,x_2)$ term since it is CP-odd.  
$F^-(x_1,x_2)=0$ for $a=g$. There are no contaminations from initial gluon 
reaction.  All contaminations come from the initial 
quark reactions which are subdominant processes at the LHC. 
There are non-zero CP violating contaminations    
only when  the observables contain  asymmetry  between $x_1$ and $x_2$.
Since the Lorentz invariant observables can always be calculated in 
the center of mass frame of the parton which depending only 
on $x_1x_2$, they will receive no contaminations. 
Therfore, within the framework of parton model, 
there are no contaminations to Lorentz invariant observables. 
In our study, they are $f_2$,$O_1$,$O_l'$,$O_2$,$O_4$.

\newpage
\vspace{1.0in}

\begin{center}
{\large Figure Captions}
\end{center}
\parindent=0pt

Fig.1 Feynman diagrams of tree level and 1-loop SUSY QCD corrections  
of CP violation to $q\bar q\to t\bar t$

Fig.2 Feynman diagrams of tree level and 1-loop SUSY QCD corrections  
of CP violation to $gg\to t\bar t$

Fig.3 Feynman diagrams of top quark CEDM  corrections  
of CP violation to $q\bar q\to t\bar t$, $gg\to t\bar t$




\eject

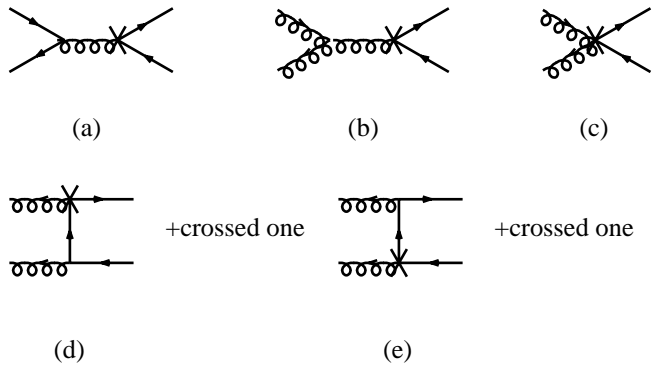
\end{document}